\documentclass[aps,physrev,preprint,groupedaddress]{revtex4-2}
\usepackage{graphicx}
\usepackage{dcolumn}
\usepackage{bm}
\usepackage{multirow}
\usepackage{array}
\usepackage{xr}
\usepackage{longtable}
\usepackage{floatrow}
\usepackage{moresize}
\usepackage{float}
\usepackage{makecell}
\usepackage{amsmath}
\usepackage{moresize}
\usepackage{relsize}
\usepackage{upgreek}

\usepackage{hyperref} 

\renewcommand{\figurename}{Figure}
\begin{document}


\title{Visualizing Poloidal Orientation in DNA Minicircles}
%

\author{Tony Lemos}
\author{Harold D. Kim}%
 \email{Contact author: harold.kim@physics.gatech.edu}
\affiliation{School of Physics, Georgia Institute of Technology\\
 Atlanta, GA 30332-0430, USA}


\date{\today}

\begin{abstract}
A short ($<$150 bp) double-stranded DNA (dsDNA) molecule ligated end-to-end forms a DNA minicircle. Due to sequence-dependent, nonuniform bending energetics, such a minicircle is predicted to adopt a certain inside-out orientation, known as the poloidal orientation. Despite theoretical and computational predictions, experimental evidence for this phenomenon has been lacking. In this study, we introduce a single-molecule approach to visualize the poloidal orientation of DNA minicircles. We constructed a set of DNA minicircles, each containing a single biotin located at a different position along one helical turn of the dsDNA, and imaged the location of biotin-bound NeutrAvidin relative to the DNA minicircle using atomic force microscopy (AFM). We applied this approach to two DNA sequences previously predicted to exhibit strongly preferred poloidal orientations. The observed relative positions of NeutrAvidin shifted between the inside and outside of the minicircle with different phases, indicating distinct poloidal orientations for the two sequences. Coarse-grained simulations revealed narrowly distributed poloidal orientations with different mean orientations for each sequence, consistent with the AFM results. Together, our findings provide experimental confirmation of preferred poloidal orientations in DNA minicircles, offering insights into the intrinsic dynamics of circular DNA.
\end{abstract}


\maketitle

\section*{Significance}
DNA minicircles are short, circular DNA molecules formed by end-to-end ligation. Their high curvature makes their conformation sensitive to the energetics of DNA bending, which varies with sequence. While computational models predict sequence-dependent poloidal orientation in minicircles, experimental evidence has been scarce. Using a novel Atomic Force Microscopy (AFM)-based assay, we demonstrate distinct poloidal orientations for different DNA sequences, providing experimental confirmation of sequence-dependent poloidal bias in DNA minicircles.

\section{Introduction}
\begin{figure}
\centering
\includegraphics[width=0.5\linewidth]{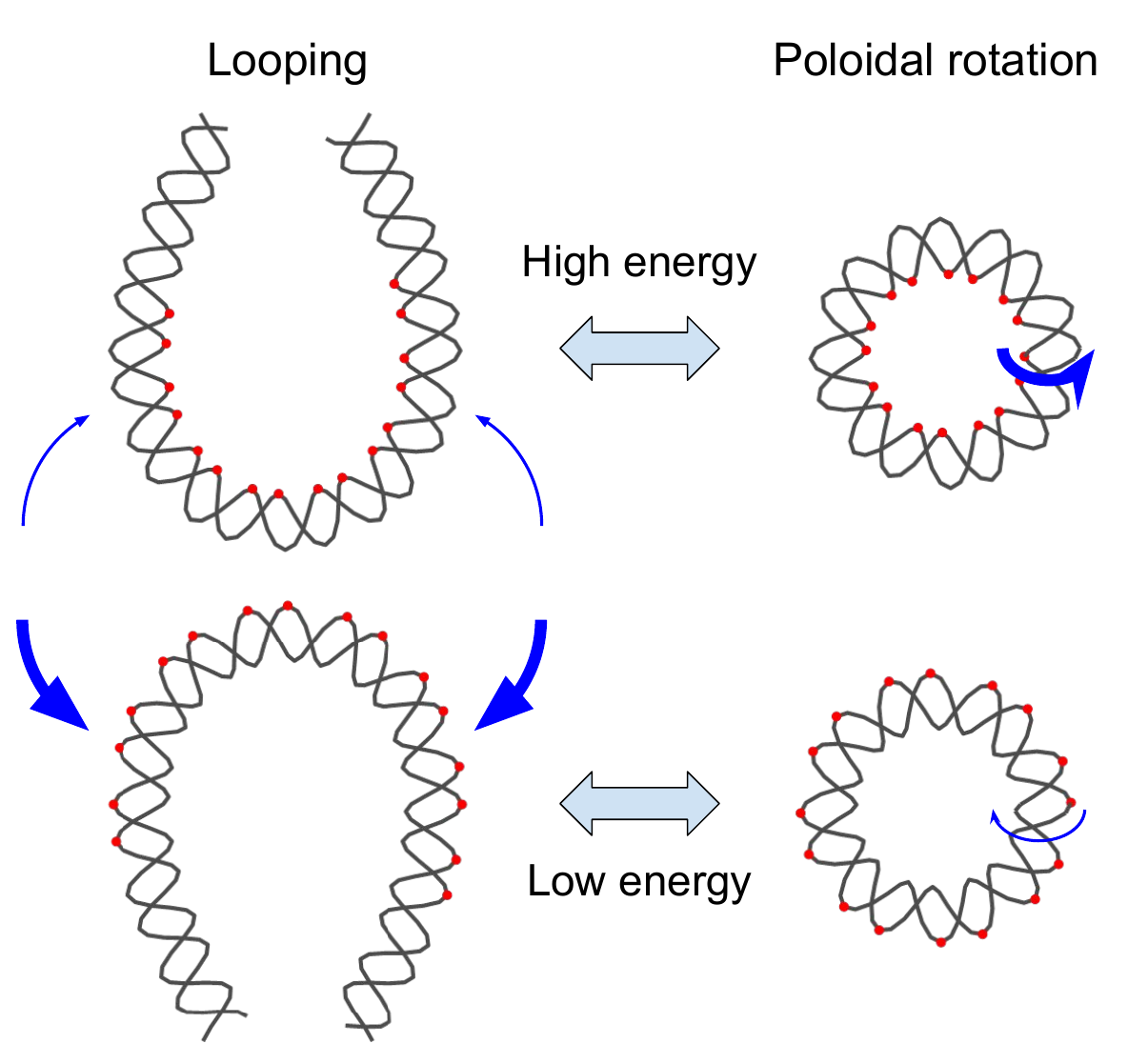}
\caption{Correspondence between DNA looping direction and poloidal orientation. For conceptual clarity, two opposite looping directions of a double-stranded DNA are illustrated. The loops are depicted in a teardrop shape, representing the minimum-energy conformation. Blue arrows indicate the direction of looping with thickness proportional to the relative frequency, and red dots mark the innermost or outermost atoms in highly bent regions. Due to sequence-dependent bending rigidity, the energetic cost of bending differs between the two directions. As a result, one looping direction is energetically favored over the other. On the right, DNA minicircles are shown with poloidal orientations corresponding to the two looped conformations. These poloidal orientations exhibit a similar energy difference, leading to a preferred orientation consistent with the favored looping direction. Blue arrows indicate poloidal rotations, and their thickness represents the relative frequency.}
\label{fig:loop-minicircle}
\end{figure}

The looping dynamics of short DNA segments ($<$100 base pairs (bp)) are largely governed by the energetics of bending at the base pair step level. Base pair step bending is highly anisotropic, energetically more favorable around the long axis (roll) than the short axis (tilt) \cite{dickerson1998dna,ma2016anisotropy}. Moreover, even among roll directions, bending toward the major groove is more favorable than bending toward the minor groove. In addition to structural anisotropy, different base pair steps exhibit varying bending rigidities due to differences in stacking interactions. As a result, the looping of short DNA preferentially follows paths that minimize the bending energy barrier, leading to a directional bias in the plane of loop formation (\autoref{fig:loop-minicircle}). Previous experimental studies suggest that such directional bias exists. One study reported that the looping rate of surface-attached short DNA changes markedly as the attachment point is shifted by several base pairs \cite{basu2021measuring}. In another study, the looping rate of a short DNA fragment oscillated with a period close to twice the helical period of DNA ($\sim$20 bp) when the DNA was symmetrically lengthened at both ends \cite{jeong2020determinants}. Both observations are consistent with a sequence-dependent directional bias in the plane of loop formation.

A more direct investigation of such bias can be achieved using DNA minicircles. In principle, a DNA minicircle can undergo inside-out rotation \cite{yoo2021dna} or twirling \cite{kulic2005twirling}. The inside-out rotational position of the minicircle can be described by the poloidal angle that goes around the helical axis of the DNA. Due to the sequence-dependent bending rigidity of base pair steps, certain poloidal angles can be more energetically favorable than others (\autoref{fig:loop-minicircle}). Indeed, all-atom molecular dynamics (MD) simulations of DNA minicircles have revealed such behavior. Whereas a minicircle of 90 bp composed of a homogeneous AT-repeat sequence constantly undergoes poloidal rotation with no preferred angle \cite{kim2022dynamics}, 85-90 bp DNA minicircles of similar length but with non-repeating sequences converge within nanoseconds toward a unimodal distribution of poloidal angles with a well-defined maximum \cite{yoo2021dna}. Notably, the most preferred poloidal angle of a minicircle derived from half of the strong nucleosome-positioning 601 sequence matches its poloidal orientation observed in the nucleosome \cite{yoo2021dna}. This computational study not only confirms sequence-dependent poloidal orientation in DNA minicircles but also identifies it as a potential driving force in nucleosome formation.

Beyond their poloidal rotation, DNA minicircles serve as an excellent model system for both experimental and computational studies of DNA mechanics. Experimentally, DNA minicircles can be readily generated by self-ligation of double-stranded DNA (dsDNA) with sticky ends. In fact, the rate of minicircle formation has long been used as a proxy for measuring looping probability density, or the J factor \cite{cloutier2004spontaneous, geggier2010sequence, du2005cyclization,basu2021measuring}. MD simulations and cryoEM imaging of DNA minicircles have provided insights into various deformation modes and their dynamics in the strong-bending regime \cite{pasi2016dna, yoo2021dna,mills2020atomistic,kim2021sequence}, which can illuminate the behavior of extrachromosomal circular DNA \cite{kumar2017normal,moller2018circular,paulsen2019small}. DNA minicircles can also be designed to exert torsional stress, enabling studies of supercoiling and twist-dependent phenomena \cite{mitchell2011atomistic,mitchell2013thermodynamics,sutthibutpong2016long,curuksu2023spectral}. Moreover, they offer a model for investigating protein binding to curved DNA. For example, when DNA is prebent into a minicircle with a defined poloidal orientation, the binding affinity of proteins can be either enhanced or inhibited \cite{khan2023deepbend,parvin1995pre,kim1997prebending}. In nanotechnology, DNA minicircles have been assembled into catenanes \cite{lohmann2014novel} and rotaxanes \cite{ackermann2010double}, enabling the creation of molecular machines \cite{valero2018bio,valero2019design,yu2021self}. They have also been explored for therapeutic applications \cite{thibault2017production}; for instance, a 95-bp DNA minicircle has been shown to function as a therapeutic agent against triple-negative breast cancer \cite{casas2022dna}. Thus, a comprehensive understanding of DNA minicircle dynamics will broaden its potential across many disciplines.

Although MD simulations have provided important insights into the poloidal angle landscape and sequence-dependent mechanics of DNA minicircles \cite{kim2021sequence,kim2022dynamics,yoo2021dna}, experimental methods to directly probe poloidal orientation have been lacking. To address this gap, we developed a single-molecule approach to visualize the poloidal orientation of DNA minicircles. We constructed a series of DNA minicircles, each containing a single biotin placed at a different position along one helical turn of the duplex, and imaged the relative location of biotin-bound NeutrAvidin using atomic force microscopy (AFM). The NeutrAvidin location alternated between the inside and outside of the minicircle in a periodic manner, and the phase of this shift differed between two sets of DNA minicircles derived from distinct sequences. Coarse-grained simulations using the mechanically accurate MADna model \cite{assenza2022accurate} further revealed narrowly distributed poloidal angles with distinct means for each sequence, consistent with our experimental observations. Together, our results provide the first direct experimental evidence of sequence-dependent poloidal bias in DNA minicircles.

\section{Materials and Methods}

\subsection{AFM Experiments}
\subsubsection{DNA Minicircle Preparation}
105-bp DNA templates were PCR-amplified from plasmid DNAs (Twist Bioscience) using primers listed in \autoref{tab:DNA_SequencesAFM}. We designed two sets of DNA templates: the first set based on the left half of the 601 nucleosome-positioning sequence \cite{lowary1998new}, and the second containing six A-tracts \cite{koo1990determination,parvin1995pre}. These two sets are referred to as 601 series and A-tract series, respectively (\autoref{sfig:sequences}). Modified primers (Eurofins Genomics and Bioneer Incorporation) were used to incorporate a biotin linker and terminal phosphate groups into the linear DNA templates (\autoref{tab:DNA_SequencesAFM}). The templates were circularized using the DNA bending protein HMG1 (Sigma Aldrich) by incubating 17.5 nM of modified linear DNA with 0.75 $\mathrm{\mu}$M HMG1 in T4 ligase buffer for 10 minutes, as described in \cite{hart2023weak}. 400 U of T4 DNA ligase (New England Biolabs) was then added and the reaction incubated overnight at 16\textdegree C. To stop the reaction, samples were heat-inactivated at 65\textdegree C for 10 minutes. Unligated linear DNA was removed by digestion with 10 U of T5 exonuclease (New England Biolabs), followed by incubation at 37\textdegree C for 30 minutes. Proteins were removed by adding 20 $\mu$g of Proteinase K (Macherey-Nagel) and incubating at 57\textdegree C for 30 minutes. The resulting DNA minicircles were purified by phenol extraction followed by PCR purification (IBI Scientific).  

\subsubsection{AFM Imaging of DNA minicircles} 
Atomic Force Microscopy (AFM) imaging of DNA minicircles bound to proteins was performed following a recently published protocol \cite{haynes2022atomic}. Approximately 30 ng of DNA minicircles were incubated with 10 ng of NeutrAvidin (Thermo Fisher Scientific) for 30 minutes at room temperature. 20 $\mu$L of magnesium absorption buffer (10 mM Tris, pH 8; 50 mM MgCl$_2$) was deposited onto a freshly cleaved muscovite mica disk (Mica Grade V-4, Structure Probe Inc.). The DNA-protein complex solution was then applied to the mica disk and incubated for 5 minutes in a humidity chamber. The mica disk was washed three times with nickel imaging buffer (20 mM HEPES, pH 8; 3 mM NiCl$_2$) using vigorous pipetting, followed by removal of the wash solution. Subsequently, 20 $\mu$L of fresh nickel imaging buffer was added, and a microfluidic chamber (MTFML-V2; Bruker) was placed over the mica for fluid-phase imaging. Imaging was conducted using a Multimode 8 AFM system (Bruker) in PeakForce Tapping mode with PEAKFORCE-HIRS-F-B cantilevers (Bruker). The cantilever was pre-equilibrated in imaging buffer for 10 minutes prior to scanning. Low-resolution, large-area scans (2-3 $\mu$M scan size; 2-3 Hz scan rate; 256x256 lines per scan) were first acquired to locate suitable scanning regions. High-resolution scans (800 nm scan size; 2.5 Hz scan rate; 512x512 lines per scan) were then collected using the following settings: 10 nm PeakForce amplitude, 4 kHz PeakForce frequency, and PeakForce setpoints ranging from 10-40 mV. A minimum of 120 images of DNA-protein complexes were collected across 2-3 independent experimental trials for each minicircle.   

\subsubsection{AFM Image Analysis} 
\begin{figure}
\centering
\includegraphics[width=\textwidth]{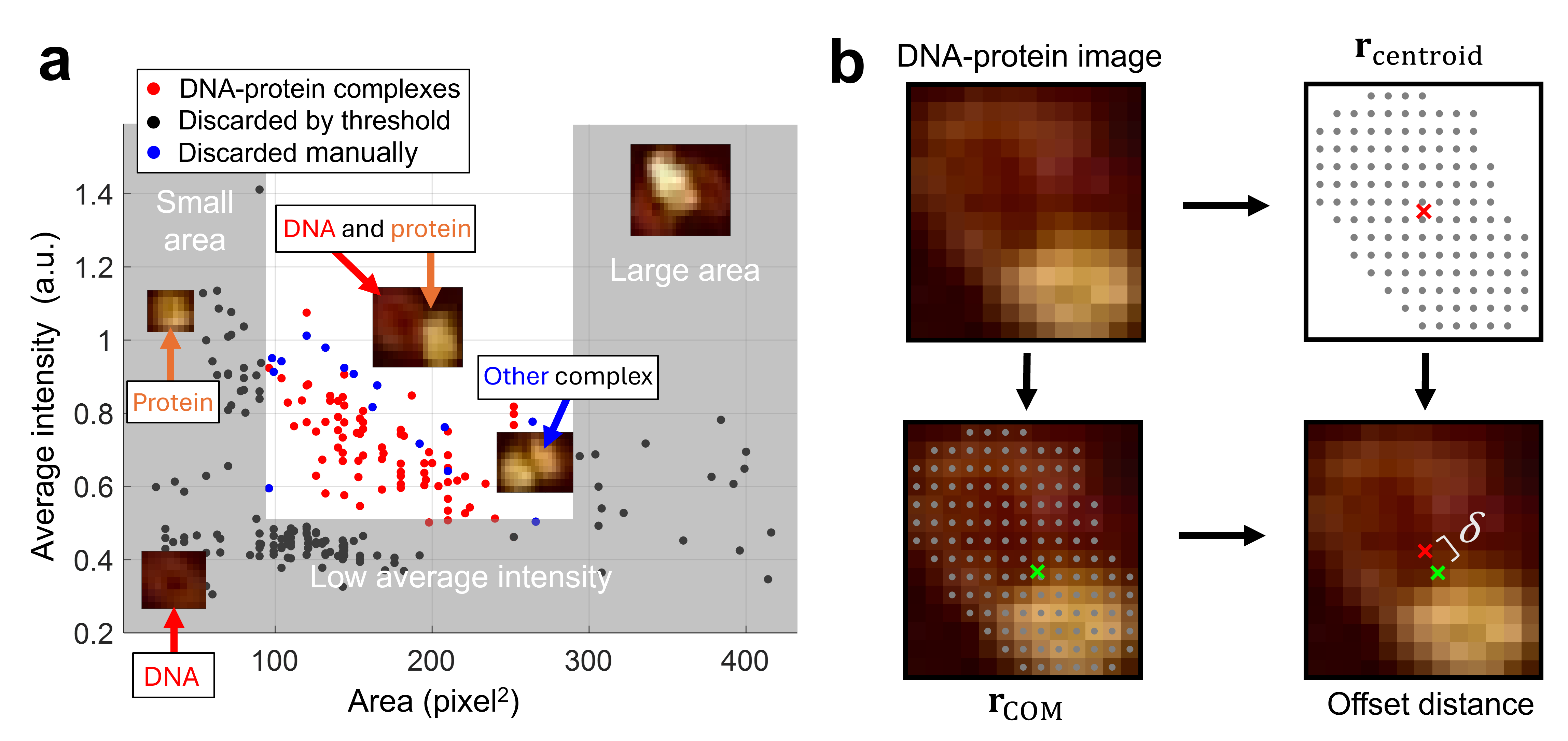}
\caption{\label{fig:ImageAnalysis} Analysis of images collected from AFM experiments.
(a) Procedure for selecting DNA–protein images from the initial collection of images. A subset of images representing lone minicircles, lone proteins, or aggregates is discarded based on area and intensity thresholds (gray boxes). The remaining images are manually inspected to distinguish DNA–protein complexes (red dots) from other types (blue dots). Representative molecules from each category are shown. (b) Calculation of offset distance $\delta$. From each selected DNA–protein image, the centroid $\mathbf{r}_\mathrm{centroid}$ (red ×) and the center of mass $\mathbf{r}_\mathrm{COM}$ (green ×) were computed based on the pixel mask of the DNA–protein complex. The offset distance $\delta$ was defined as the Euclidean distance between $\mathbf{r}_\mathrm{centroid}$ and $\mathbf{r}_\mathrm{COM}$.}
\end{figure}

AFM images were processed using NanoScope Analysis 3.00 (Bruker). Second-order image flatting was applied to correct for tilt and bow artifacts, and a standard color bar scale ranging from –2 nm to 6 nm was used to obtain good contrast between the molecules and the surface. Image analysis was performed using ImageJ \cite{schneider2012nih}. An intensity threshold was first applied to select molecules within the AFM image. Bounding rectangles were then used to isolate and extract molecules to form an initial collection of images. Image selection for DNA-protein complexes was performed by applying thresholds based on average pixel intensity and image area (\autoref{fig:ImageAnalysis}), thereby removing molecules that were too large (DNA-protein aggregates), too small (protein only), or too dim (DNA only). The remaining images were manually inspected to confirm the presence of DNA-protein complexes. For each validated DNA-protein image (\autoref{sfig:AFMscansSupp}), a solid, connected polygonal mask was generated from the initial intensity threshold-clearing pixels. The geometric center (centroid) $\mathbf{r}_\mathrm{centroid}$ was computed as the arithmetic mean of the coordinates of all pixels comprising the mask. The center of mass $\mathbf{r}_\mathrm{COM}$ was also calculated using the same pixel coordinates, with weights assigned based on pixel intensity (\autoref{fig:ImageAnalysis}). The offset distance $\delta$ was then calculated as the Euclidean distance between $\mathbf{r}_\mathrm{centroid}$ and $\mathbf{r}_\mathrm{COM}$, providing a measure of the protein's position relative to the entire DNA-protein complex. 

\subsection{Course-grained Simulations}
We performed MADna simulations \cite{assenza2022accurate} using LAMMPS \cite{LAMMPS} on a laptop computer. We chose MADna because it more accurately captures sequence-dependent DNA mechanics compared to other coarse-grained models \cite{assenza2022accurate}, and atomistic simulations are too computationally expensive for the scope of this study. In the MADna model, each nucleotide is coarse-grained to three atoms, sugar (S), phosphate (P), and base (B). These atoms are connected via two-, three-, and four-body bonded interactions, referred to as bonds, angles, and dihedrals, respectively. Simulations were run for 100 ns with a time step of 20 ps, in 150 mM salt at room temperature (25°C). For all tested sequences, the poloidal angle converged to its equilibrium distribution within the first 5 ns. Therefore, we excluded the initial 5 ns from the equilibrium analysis.

\subsubsection{Minicircle construction}
To create a DNA minicircle for MADna simulations, we first constructed an ideal linear DNA molecule where each atom type follows a perfect helical trajectory. Each of the six helical trajectories ($i=1,\ldots,6$ for three atoms and two strands) requires the base pair step twist ($\Delta\theta$), the base pair step rise ($\Delta z$), radius ($r$), angular offset ($\theta_0$), and rise offset ($z_0$):
\begin{align*}
x_{i,n} &= r_i \cos (n\Delta \theta +\theta_{i,0}) \\
y_{i,n} &= r_i \sin (n\Delta \theta +\theta_{i,0}) \\
z_{i,n} &= n\Delta z+z_{i,0}
\end{align*}
Here $i$ represents any one of six atom types (sugar, base, and phosphate from either strand), and $n$ is the base pair index. We analyzed a 90-bp linear DNA molecule with a random sequence, initialized by MADna, to obtain these parameters. First, the sugar helix parameters were determined from the average sugar-sugar distance, helical twist, and helical rise. Next, using a reference frame defined by the sugar atoms in three successive base pairs ($n-1$, $n$, and $n+1$), we determined the relative positions of all other atoms in the same base pair and averaged them over all $n$ to obtain all remaining helix parameters. Finally, we adjusted $\Delta \theta$ to ensure uniformly twisted minicircle helices, setting it to $\Delta \theta = 360^\circ/105$ for a 105-bp minicircle, for example. These parameters, used to model the linear DNA, are summarized in \autoref{tab:DNAinitial}.

We then transformed the linear DNA aligned along the z-axis to a minicircle following the published method \cite{yoo2021dna}. First, the molecule was translated so that its center of mass is located at $(0,R,0)$. Each atom was transformed according to $(x,y,z)\rightarrow(x,y\cos(z/y),y\sin(z/y))$. This transformation produces a uniformly twisted planar minicircle. In addition to the position coordinates that describe a minicircle, new bonds, angles, and dihedrals must be added to ensure a contiguous double-stranded structure for MADna simulations. We added six bonds (SP, PS, and BB-intra for each strand), six angles (SPS, 3PSB5, and 5PSB3 for each strand), and 16 dihedrals (two SPSP, two PSPS, one SPSB53, one SPSB35, one PSBB53, and one PSBB35 for each strand) between the end atoms. 

\subsubsection{Poloidal Angle Calculations}
We define the poloidal angle based on the global morphology of the DNA minicircle, which differs slightly from previous definitions \cite{yoo2021dna}. Our approach yields a quantity that is less sensitive to local distortions of the helix. As shown in \autoref{sfig:PoloidalCalc}, we select four equidistant phosphate atoms: $\mathrm{P}_0$, $\mathrm{P}_{1/4}$, $\mathrm{P}_{1/2}$, and $\mathrm{P}_{3/4}$. We then define x and z axes along two vectors connecting the pairs ($\mathrm{P}_0$, $\mathrm{P}_{1/2}$) and ($\mathrm{P}_{1/4}$, $\mathrm{P}_{3/4}$), respectively. The cross product of the two vectors yields the y-axis.

Next, we compute the geometric center (centroid) of a 10-bp DNA segment containing an equal number of atoms (S, P and B) on each side of $\mathrm{P}_0$ and find its projection on the x–y plane; this point is designated as the origin O. In this construction, both $\mathrm{P}_0$ and O lie in the x–y plane, and the polar angle of $\mathrm{P}_0$ in this plane defines the poloidal angle. This angle can be extracted from the x- and y-components of $\mathrm{P}_0$ using the two-argument arctangent function, atan2. For plotting mean-centered histograms of poloidal angles (\autoref{sfig:PoloidalCalc}), we rotate all position vectors by the negative mean of the poloidal angles.  

\section{Results}

We sought to determine whether DNA minicircles exhibit a preferred poloidal orientation, or poloidal bias at the single molecule level. Direct visualization of a poloidal bias would require localizing a specific atom relative to the helical axis of the circular DNA. However, no current imaging technique provides sufficient resolution to achieve this. Therefore, we employed a secondary protein as a marker of poloidal orientation.

\begin{figure}
\centering
\includegraphics[width=\textwidth]{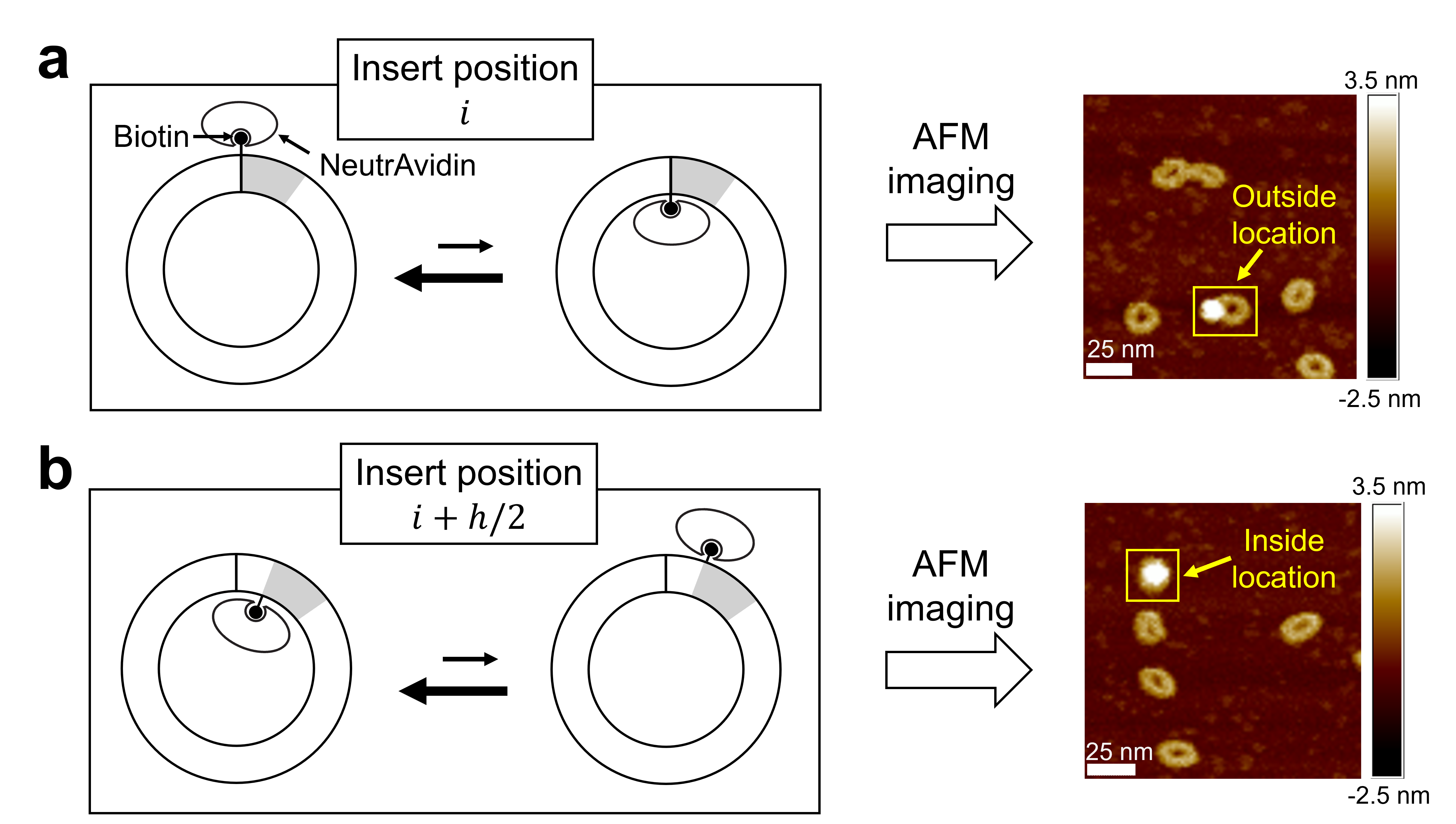}
\caption{\label{fig:AFMexperiment} Schematic of poloidal bias visualization using AFM. (a) A DNA minicircle with a biotin-containing segment (gray) inserted at position $i$. Suppose the DNA minicircle has a preferred poloidal orientation, which corresponds to an outward orientation of the biotin. Due to this outward orientation, biotin-bound NeutrAvidin will appear at the outside location, as shown by the AFM image to the right.
(b) A DNA minicircle, identical to the previous one except that the biotin segment is shifted by half a helical period to position $i + h/2$. Here, the biotin is more likely to face inward due to the preferred poloidal orientation of the DNA minicircle. This will result in NeutrAvidin appearing at the inside location, as shown by the AFM image to the right.}
\end{figure}

To implement this strategy, we incorporate a single biotin at a defined position within the DNA minicircle and use the NeutrAvidin protein as a marker. The resulting DNA–protein complexes are deposited and imaged using AFM. If the DNA minicircle lacks poloidal bias, the biotin-bound marker should appear on the inside and outside of the minicircle with equal probability. Conversely, if a poloidal bias exists, the relative probabilities of the marker appearing on the inside versus the outside should differ (\autoref{fig:AFMexperiment}). This inside-outside asymmetry should also vary systematically as the biotin position is incrementally shifted over one helical turn of DNA ($h$). For example, if the marker is observed on the outside of the minicircle when the biotin is at position $i$, it should appear on the inside when the biotin is placed at position $i+h/2$ (\autoref{fig:AFMexperiment}).

\subsection{Poloidal bias detection using AFM}

\begin{figure}
\centering
\includegraphics[width=\textwidth]{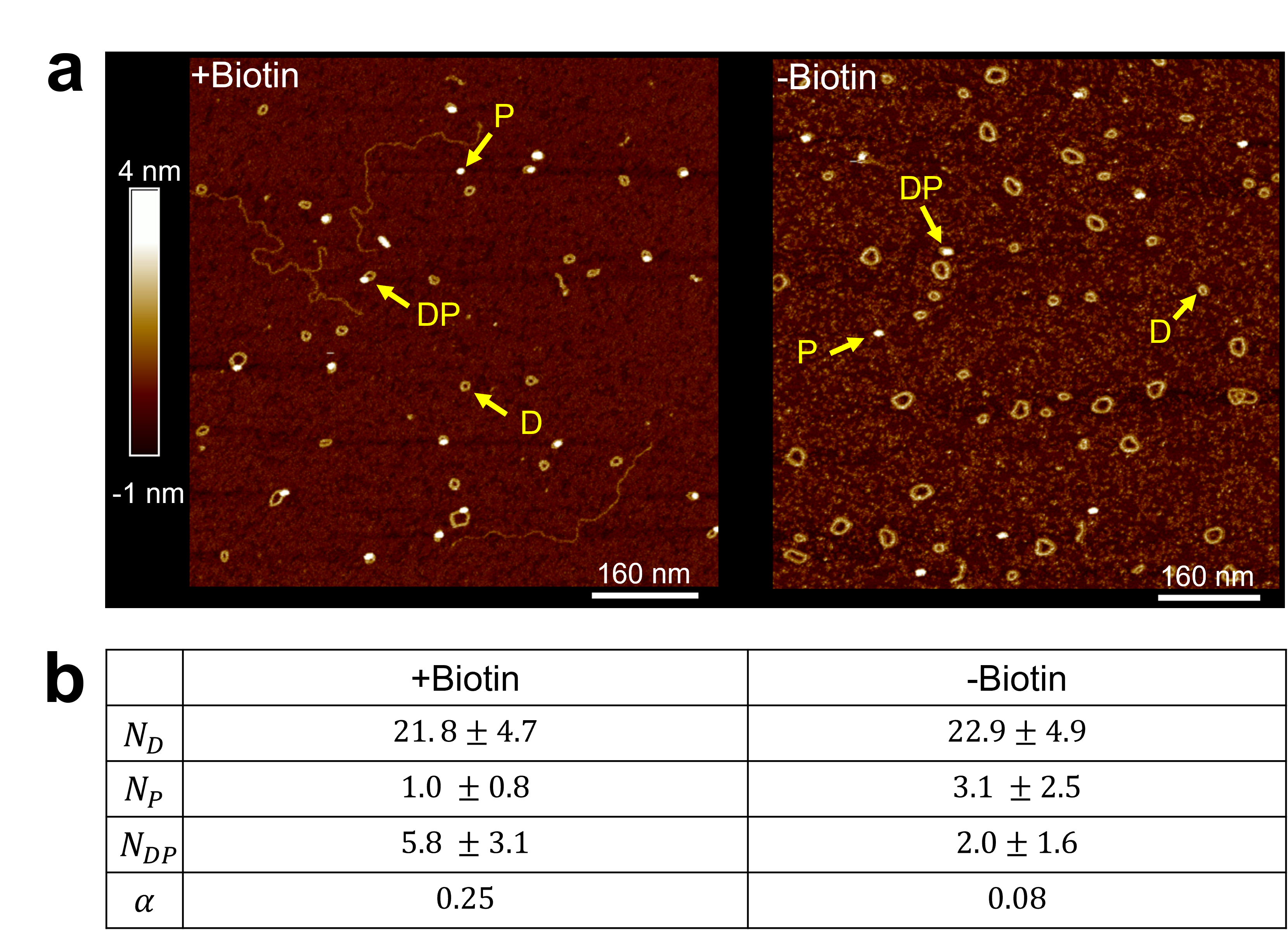}
\caption{\label{fig:AFMCtrlexperiment} Colocalization of DNA minicircle and NeutrAvidin with and without biotin. (a) An AFM image of biotinylated DNA minicircles (+Biotin) incubated with NeutrAvidin is shown on the left, and a corresponding image of non-biotinylated DNA minicircles (–Biotin) is shown on the right. Arrows indicate examples of DNA (D), protein (P), and DNA–protein complexes (DP). Larger loops in the AFM images represent circular dimers, which are by-products of the ligation protocol and are excluded from analysis.  
(b) Number statistics of different molecule types per scan area. The table shows the mean and standard deviation for each molecule type observed in +Biotin and –Biotin samples. The calculated relative affinity ($\alpha$) values are also listed.}
\end{figure}

We first performed control experiments to verify that the observed DNA-protein complexes resulted from specific binding of NeutrAvidin to DNA minicircles via biotin, rather than from random associations. To this end, we constructed two DNA minicircles with identical sequences (601\_6 in \autoref{tab:DNA_SequencesAFM}): one containing a biotin and one without. \autoref{fig:AFMCtrlexperiment} shows representative AFM images of the two minicircles incubated with equal concentrations of NeutrAvidin. We observed a higher population of DNA–protein complexes for the biotin-modified minicircle compared to the biotin-free version, indicating that biotin promotes NeutrAvidin binding. We also observed a smaller, but non-negligible, number of DNA–protein complexes in the absence of biotin, suggesting that NeutrAvidin can associate with the DNA minicircle through nonspecific interactions.

To quantitatively compare specific versus nonspecific binding, we calculated the relative affinity 
\begin{equation}
\alpha=\frac{N_{DP}}{N_D N_P}   
\end{equation}
where $N_D$, $N_P$, and $N_{DP}$ represent the number of DNA minicircles, protein molecules, and DNA-protein complexes, respectively, identified in an AFM scan. Assuming that the number of surface-bound molecules reflects their corresponding solution-phase concentrations, this quantity is proportional to the association constant. The relative affinity $\alpha$ for both minicircle types is shown in \autoref{fig:AFMCtrlexperiment}. As expected, the biotin-modified minicircle exhibited an affinity approximately 3.3-fold higher than that of the unmodified minicircle. This result indicates that more than 75\% of the observed DNA–protein complexes arise from biotin-specific interactions. Therefore, the relative position of NeutrAvidin in these complexes can be used to detect the poloidal bias in DNA minicircles. 

To quantify the relative position of the marker within the DNA–protein complex, we calculated the offset distance $\delta$ between the centroid of the complex and the center of intensity of the bright, blob-like region identified as the putative DNA–protein complex in AFM images. The value of $\delta$ is expected to be small when the marker is located inside the DNA minicircle, and to increase when the marker lies on the outside. Therefore, for a DNA minicircle with a strong poloidal bias, we expect $\delta$ to oscillate as the biotin position is incrementally shifted over one helical turn. To this end, we constructed a series of 105-bp DNA minicircles that shared a common sequence derived from the left half of the 601 sequence \cite{lowary1998new}, but with the biotin-modified segment inserted at seven different positions in 2-bp increments, labeled from 0 to 12 (\autoref{sfig:sequences}). To achieve this positional variation, we inserted a 20-nt segment containing a biotin-modified deoxythymidine (dT) into different positions within the 105-nt core sequence, allowing for precise and consistent control of biotin positioning along one strand.

\begin{figure}
\centering
\includegraphics[width=\textwidth]{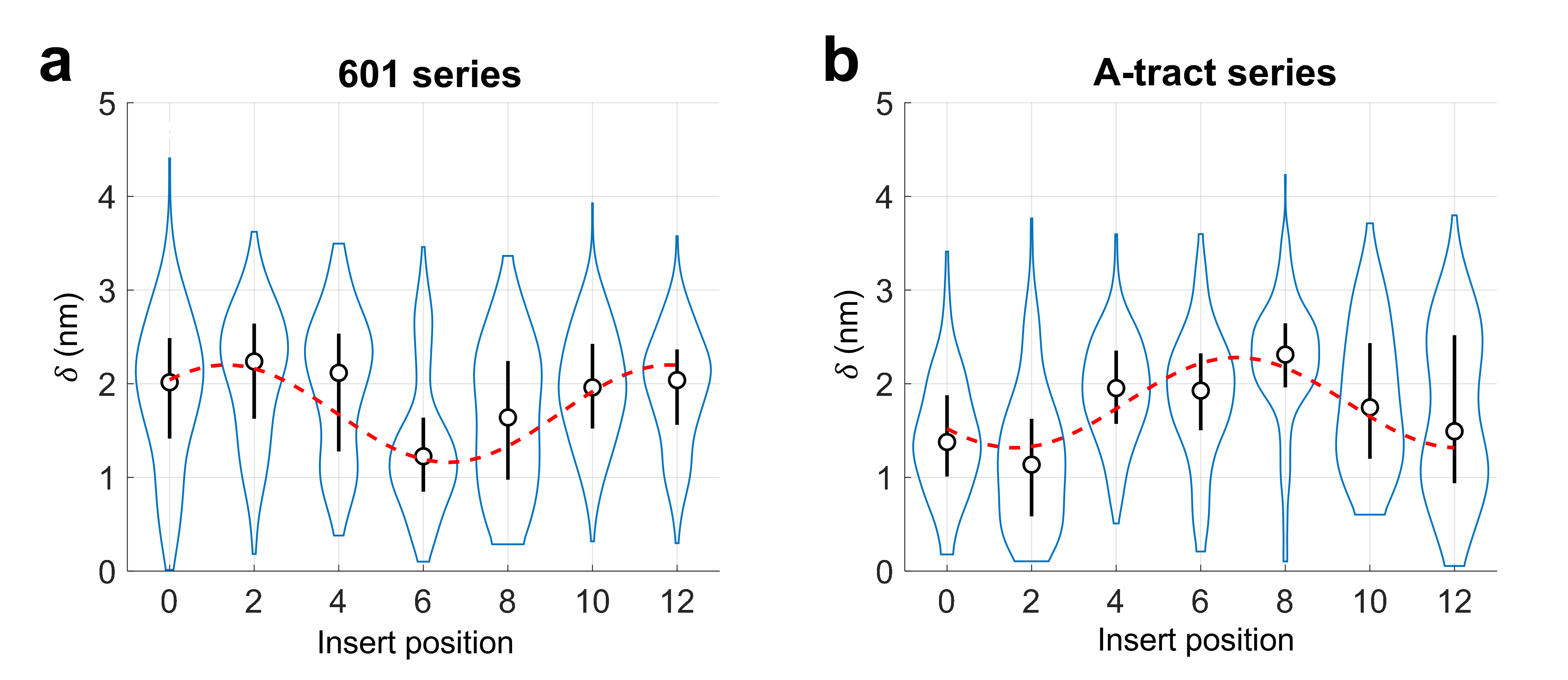} 
\caption{\label{fig:AFMScanResults}Offset distances $\delta$ extracted from AFM images of DNA-protein complexes.
The distribution of $\delta$ is shown as a violin plot. The circle represents the median, and the vertical line indicates the standard deviation.
(a) Offset distances from DNA minicircles of the 601 series. A sinusoidal fit (red dashed line) to the plot yields a peak at insert position 1.3.
(b) Offset distances from DNA minicircles of the A-tract series. A sinusoidal fit to the plot yields a peak at insert position 6.8.}
\end{figure}

The extracted values of $\delta$ for the 601 series of minicircles are shown in \autoref{fig:AFMScanResults}. For insert position 0 (601\_0), $\delta$ was approximately 2 nm, close to the apparent radius of the DNA minicircle, indicating that NeutrAvidin is positioned far from the center. At insert position 2 (601\_2), $\delta$ increased slightly, with a distribution similar to that of position 0, suggesting further outward displacement. As the biotin was shifted further, $\delta$ decreased, reaching a minimum at position 6. AFM images for insert position 6 showed a more centralized intensity pattern, with a median $\delta$ of approximately 1.2 nm. Beyond position 6, $\delta$ increased again, returning to near-maximal values ($\sim2$ nm) at positions 10 and 12 (\autoref{fig:AFMScanResults}). This oscillatory change in $\delta$, from high to low to high with a periodicity matching that of the DNA helix, is consistent with the presence of poloidal bias.

To test whether the observed poloidal bias was dictated by the core sequence of the DNA minicircle, we selected an alternative sequence containing six A-tracts, which is known to exhibit strong poloidal bias due to its intrinsic curvature \cite{koo1990determination}. Indeed, such a DNA minicircle was previously used to control the curvature direction of protein-binding sites \cite{parvin1995pre}.
We constructed this A-tract series of minicircles by inserting the same 20-nt biotin-modified segment into the six A-tract core sequence at the same set of positions, labeled 0 through 12 (\autoref{sfig:sequences}). As shown in \autoref{fig:AFMScanResults}, the extracted $\delta$ values from the A-tract series display an oscillatory pattern similar to that of the 601 series, but starting from a low $\delta$, reflecting a clear phase shift from that for the 601 series. To quantify this phase shift, each oscillatory curve was fitted by a sinusoidal function weighted by the standard deviation. The A-tract curve lags the 601 curve by 5.5 bp, which corresponds to a 189\textdegree\space phase shift. This clear phase shift in $\delta$ between the two minicircle series strongly supports the idea that the preferred poloidal orientation in DNA minicircles is sequence-dependent.

\subsection{Poloidal bias in course-grained simulations}

\begin{figure}
\centering
\includegraphics[width=\textwidth]{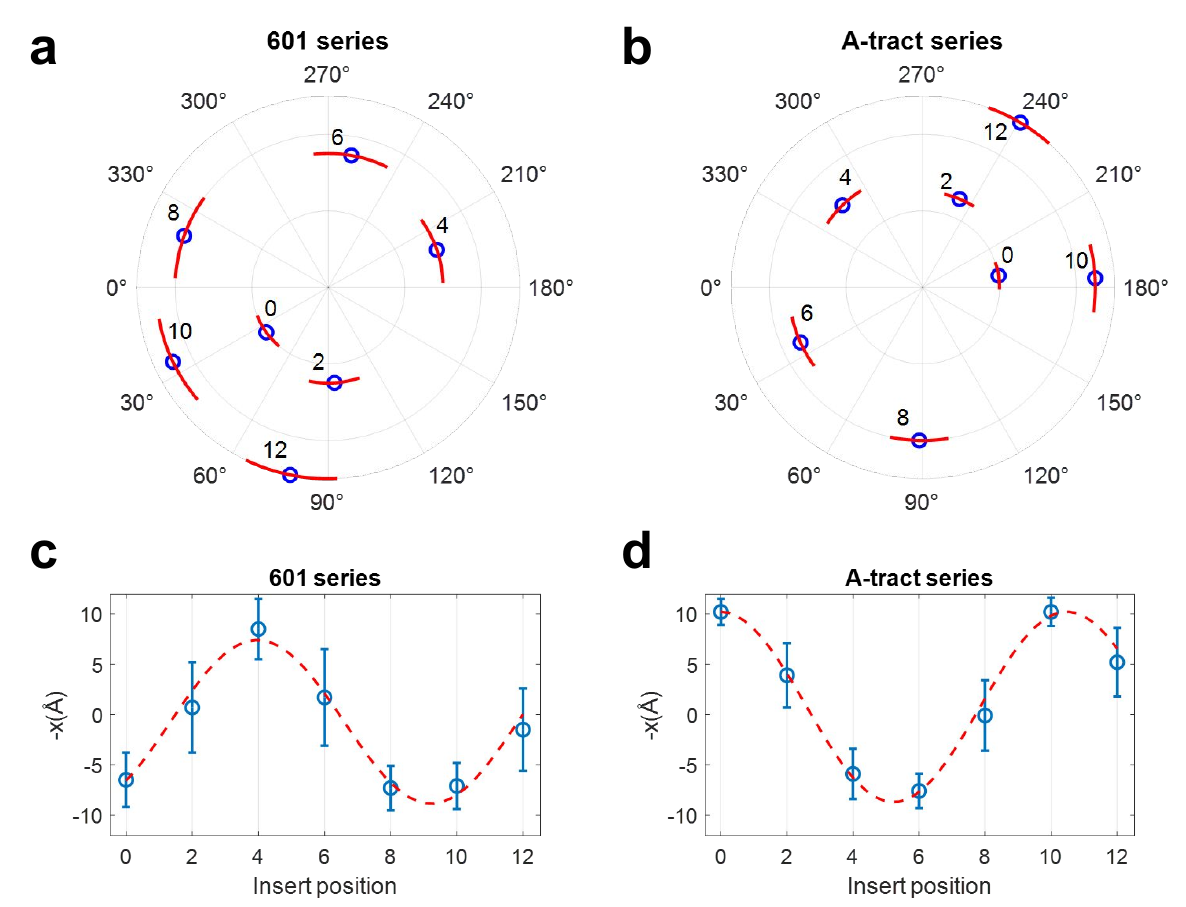}
\caption{\label{fig:SimResults} Poloidal-angle statistics of DNA minicircles obtained from MADna simulations. The statistics were computed based on the phosphate atom closest to the biotin-linked dT. (a) Polar plot of the mean and standard deviation of poloidal angles for 601 series. Blue circles represent the mean angles, while red arcs indicate their standard deviations. The number adjacent to each point indicates the insert position. Data points for different insert positions are radially distributed for visual clarity. (b)  Polar plot of the mean and standard deviation of poloidal angles for minicircles in the A-tract series. (c) Mean and standard deviation of the inverted horizontal position ($-x$) of the phosphate atom for the 601-series minicircles. $x$ is inverted to trend in the same direction as $\delta$ extracted from AFM images. Blue circles represent the horizontal displacement relative to the helical axis (the origin in the polar plot), and error bars indicate standard deviations. The red dashed curve shows a sinusoidal fit to the means, weighted by their errors, with a fixed period of 10.5. The fit curve has a peak at insert position 4. (d) Same as (c), but for the A-tract series minicircles. The fit curve has a peak at insert position 10.5.}
\end{figure}

Encouraged by the ability of our AFM-based assay to detect sequence-dependent poloidal bias in DNA minicircles, we sought an independent approach to further validate the presence of poloidal bias. One promising approach is to use all-atom MD simulations, which have successfully captured poloidal orientations in DNA minicircles \cite{yoo2021dna,kim2022dynamics}. However, all-atom simulations are computationally intensive, and our AFM-based assay does not provide atomic-level precision for meaningful comparison. In this regard, coarse-grained DNA simulations offer a more practical alternative. Yoo et al. \cite{yoo2021dna} attempted to investigate the poloidal angle landscape of DNA minicircles using coarse-grained models such as oxDNA2 \cite{snodin2015introducing} and 3SPN.2C \cite{hinckley2013experimentally}, but failed to obtain a stable equilibrium. Instead, the simulated minicircles frequently formed kinks or transiently unraveled.

We therefore chose to use the recently developed MADna model to simulate DNA minicircles \cite{assenza2022accurate}. MADna is a three-site-per-nucleotide coarse-grained model that includes up to three-body bonded interactions and is parametrized using all-atom MD simulations. It can accurately reproduce the phenomenon of DNA overwinding upon stretching that other models fail to, and makes improved predictions for elastic constants, twist–bend coupling, and the sequence dependence of helical pitch and persistence length. More importantly, MADna does not permit bond-breaking events such as kink formation or local melting, and thus avoids the stability issues encountered with oxDNA2 and 3SPN.2C in minicircle simulations \cite{yoo2021dna}.

We performed MADna simulations on all 14 DNA minicircles from two sequence series: 601 and A-tract. Each minicircle was initialized in a perfectly circular conformation and simulated for 100 ns in 150 mM salt at 25\textdegree C. The poloidal angle was defined using the position of a phosphate atom \cite{yoo2021dna}, which represents the outermost atom of the coarse-grained DNA model. Specifically, we selected the phosphate atom nearest to the biotin-modified dT used in our AFM assay, with the expectation that this atom would approximate the location of the bound NeutrAvidin. However, as illustrated in \autoref{sfig:BiotinDNA}, the precise relationship between the poloidal angle defined by this phosphate and the actual position of the biotin or NeutrAvidin remains uncertain due to the length and flexibility of the biotin linker. 

The poloidal angle distribution reached equilibrium within the first 5 ns; therefore, we used the remaining 95 ns of each trajectory to compute equilibrium statistics. The phosphate atom’s position relative to the helical axis is shown in \autoref{sfig:PoloidalCalc}, and the corresponding poloidal angle distributions are shown in \autoref{sfig:PoloidalCalc}. Circular statistics were used to compute the mean and standard deviation of the poloidal angle for each DNA minicircle (see \autoref{tab:angle_x_stats}). As shown in \autoref{fig:SimResults}, the poloidal angles for all minicircles were confined to a narrow range, indicating that the MADna model successfully captured their poloidal bias. As the position of the biotin-insert was shifted from position 0 to 12, the mean poloidal angle rotated by nearly 400°, while the standard deviation remained relatively constant. This suggests that the 20-bp biotin insert follows the poloidal orientation imposed by the 85-bp core sequence, rather than influencing it.

Most importantly, the simulations show that DNA minicircles with the same insert position but different core sequences exhibit distinct mean poloidal angles. For example, at insert position 0, the mean poloidal angle is 35.9° for the 601 series and 189° for the A-tract series. This result confirms the presence of sequence-dependent poloidal bias in DNA minicircles. To quantitatively compare the poloidal bias between the two series, we computed the statistics of the horizontal displacement $x$ of the phosphate atom relative to the helical axis of DNA (See \autoref{sfig:PoloidalCalc}). A positive $x$ indicates a position inside the minicircle, whereas a negative $x$ corresponds to a position outside the minicircle. As shown in \autoref{fig:SimResults}, the phosphate position oscillates between the inside and outside of the minicircle as the insert position changes. Based on sinusoidal fits, the A-tract curve lags the 601 curve by 6.5 bp. Although this value is slightly larger than 5.5 bp  extracted in the insert-position dependence of $\delta$ (\autoref{fig:AFMScanResults}), the two values are in good agreement given the uncertainty in the fitting parameters.

We also observed notable differences in the dispersion of the poloidal angle distributions between the two series. The standard deviation for the 601 series was consistently higher than that of the A-tract series (\autoref{sfig:histograms} and \autoref{tab:angle_x_stats}). Additionally, while all distributions were unimodal, some exhibited more pronounced asymmetry (e.g., 601\_6 and A-tract\_10) than others. These results suggest that the width of the poloidal angle distribution is determined by the 85-nt core sequence, whereas the symmetry of the distribution can be influenced by local sequence variations arising from different insert positions.   

\section{Discussion}
Using AFM, we visualized the poloidal orientation of DNA minicircles by imaging the location of NeutrAvidin bound to the minicircle. We observed a clear, oscillatory shift in NeutrAvidin’s location as the biotin was incrementally moved along one helical turn relative to the core sequence, indicating the presence of a preferred poloidal orientation. Notably, the NeutrAvidin location differed between minicircles with different core sequences we tested, even when the biotin insert position was fixed relative to the core. This result confirms that poloidal orientation is sequence-dependent. To further support these findings, we employed MADna, a recently developed coarse-grained model of DNA and found a similar difference in poloidal orientation between the two core sequences. Hence, our study experimentally confirms the presence of preferred poloidal orientations in DNA minicircles predicted by previous computational studies \cite{yoo2021dna,kim2021sequence,kim2022dynamics}.

The MADna model, which preserves the double-helical structure of DNA, predicts narrow poloidal angle distributions in DNA minicircles for both sequences tested. The standard deviations range from 18° to 35°. Furthermore, MADna predicts that the mean poloidal angles or preferred poloidal orientations of the two sequences are approximately 140° out of phase. This phase difference aligns well with that in the offset distances ($\delta$) experimentally measured by our AFM assay.

However, the distribution of offset distances appears considerably broader than that of the poloidal angle, and in some cases, even displays bimodal features (e.g., 601\_4 and A-tract\_12 in \autoref{fig:AFMScanResults}). This broadening likely arises from the length and flexibility of the six-carbon linker, which reduces the correlation between the position of biotin (as marked by NeutrAvidin) and the underlying DNA. We further speculate that the apparent bimodality may result from interactions between the DNA–protein complex and the surface. Such interactions could flatten the complex, effectively bifurcating the 3D conformations into two preferred NeutrAvidin positions relative to the minicircle. A more comprehensive understanding of these effects may be obtained through all-atom MD simulations of the DNA–NeutrAvidin complex, which we intend to pursue in future studies.



In addition to the two core sequences used in this study, MADna produces sequence-dependent poloidal angle distributions that are consistent with symmetry-based predictions. Homopolymeric and dipolymeric sequences such as poly-A and poly-(GC) exhibit inherent rotational symmetry: a perfect circular DNA of poly-A sequence is invariant under a poloidal rotation by $ 360^\circ/h\approx34.3^\circ$, and poly-(GC) is invariant under rotation by $2\times360^\circ/h\approx68.6^\circ$. Consequently, the poloidal angle distribution of the corresponding DNA minicircles, if present, would be expected to repeat every $34.3^\circ$ or $68.6^\circ$, respectively. These successive distributions would substantially overlap, effectively producing a uniform distribution. MADna simulations of mononucleotide and dinucleotide repeat sequences confirm this prediction (see \autoref{sfig:poloidaldistribution}).

Based on the symmetry principle, it is also possible to program certain features in the poloidal angle landscape of a DNA minicircle. Repeat sequences with a periodicity that is in perfect synchrony with the helical period can most strongly influence the poloidal landscape, as each repeating unit contributes additively. As a result, the distribution can become significantly sharper compared to that of a random sequence. One example is shown in \autoref{sfig:poloidaldistribution}, where a repeat of \texttt{ACCCCCCCCCC} produces a more narrow distribution with a standard deviation of $14.8^\circ$, compared to other sequences (\autoref{sfig:histograms}). Such narrow distributions can be combined without overlap by inserting additional adenines at well-separated positions. For instance, a repeat of \texttt{ACCCCACCCCC} yields a bimodal poloidal angle distribution (\autoref{sfig:poloidaldistribution}), with the system switching between the two peaks several times within a 100-ns simulation. This kind of bistable or switchable behavior could be interesting in the context of interlocked DNA \cite{valero2019design}, although synthesizing DNA minicircles with arbitrary repeat sequences remains a technical challenge.

The asymmetry of the poloidal angle distribution is another intriguing feature that can be exploited. The Schiessel group proposed that a DNA minicircle can undergo continuous unidirectional twirling when externally driven by a periodically changing environment, such as temperature \cite{kulic2005dna,kulic2005twirling}. Coupled with a linear track, this unidirectional twirling could be harnessed for directed translation \cite{song2024structure}. A key requirement for the operation of this molecular ratchet is an asymmetric potential energy landscape along the poloidal direction. Our simulations show that some sequences produce more pronounced asymmetry than others. Therefore, identifying sequence features that govern the asymmetry of the poloidal angle landscape could be an interesting direction for future research with potential practical applications.

The poloidal bias studied here may represent an important mechanical property of DNA, especially for loop formation in highly compacted and constrained genomic environments. A recent study shows that the intrinsic bending direction of short DNA segments matches conformations seen in protein–DNA complexes \cite{park2024probing}. It may also bias the earliest steps of loop extrusion by SMC proteins \cite{higashi2021brownian,dekker2023molecular}. For example, SMC complexes may begin loop extrusion by capturing a small “seed” loop, whose rate of formation and direction are determined by short-scale DNA mechanics \cite{biswas2023impact}. Once initiated, loop growth requires binding of the hinge domain to a distal DNA site \cite{bonato2025spontaneously}, which would be influenced by anisotropic looping of the bridging DNA. In this context, the preferred bending direction of DNA over short length scales may be a critical determinant of the architecture of SMC-extruded loops. Future research focusing on the initial stages of SMC loop extrusion could shed light on this conjecture.

\section*{Author Acknowledgments}
 T.L. and H.D.K designed the study. T.L. performed AFM experiments. H.D.K. performed simulations. T.L. and H.D.K. contributed analytic tools, analyzed data and wrote the manuscript.

 \section*{Declaration of Interest}
 The authors declare no competing interests.

 \section*{Supporting References}
 References \cite{bergwerf2015molview} and \cite{hanwell2012avogadro} appear in the supporting material.
\appendix



\begin{acknowledgments}
We wish to acknowledge the support of the members of the Kim lab. This work was supported by the National Institutes of Health
(R01GM112882).
\end{acknowledgments}
\bibliography{references}


\clearpage
\appendix
\renewcommand{\thesection}{\Alph{section}}
\renewcommand{\thetable}{S\arabic{table}}

\newcounter{sfigure}
\newenvironment{sfigure}[1][]{
    \refstepcounter{sfigure}
    \begin{figure}[#1]
    \renewcommand{\figurename}{Figure}
    \renewcommand{\thefigure}{S\arabic{sfigure}}
}{
    \end{figure}
}

\begin{sfigure}[p]
\centering
\includegraphics[width=\textwidth]{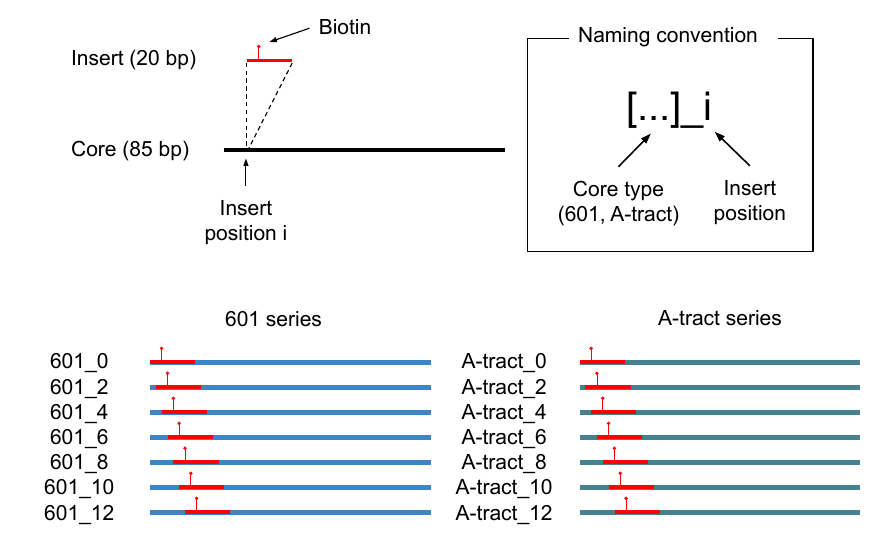}
\caption{Sequences used for DNA minicircles in this study. The 85-nt core sequence was derived from either the left half of the 601 nucleosome positioning sequence (601 series) or from six in-phase A-tracts (A-tract series). A 20-nt segment containing a biotin-modified deoxythymidine was inserted at seven different positions. The sequences are named based on the core sequence and the insert position, and are listed in \autoref{tab:DNA_SequencesAFM}.}
\label{sfig:sequences}
\end{sfigure}
\clearpage

\begin{sfigure}[p]
\centering
\includegraphics[width=0.8\textwidth]{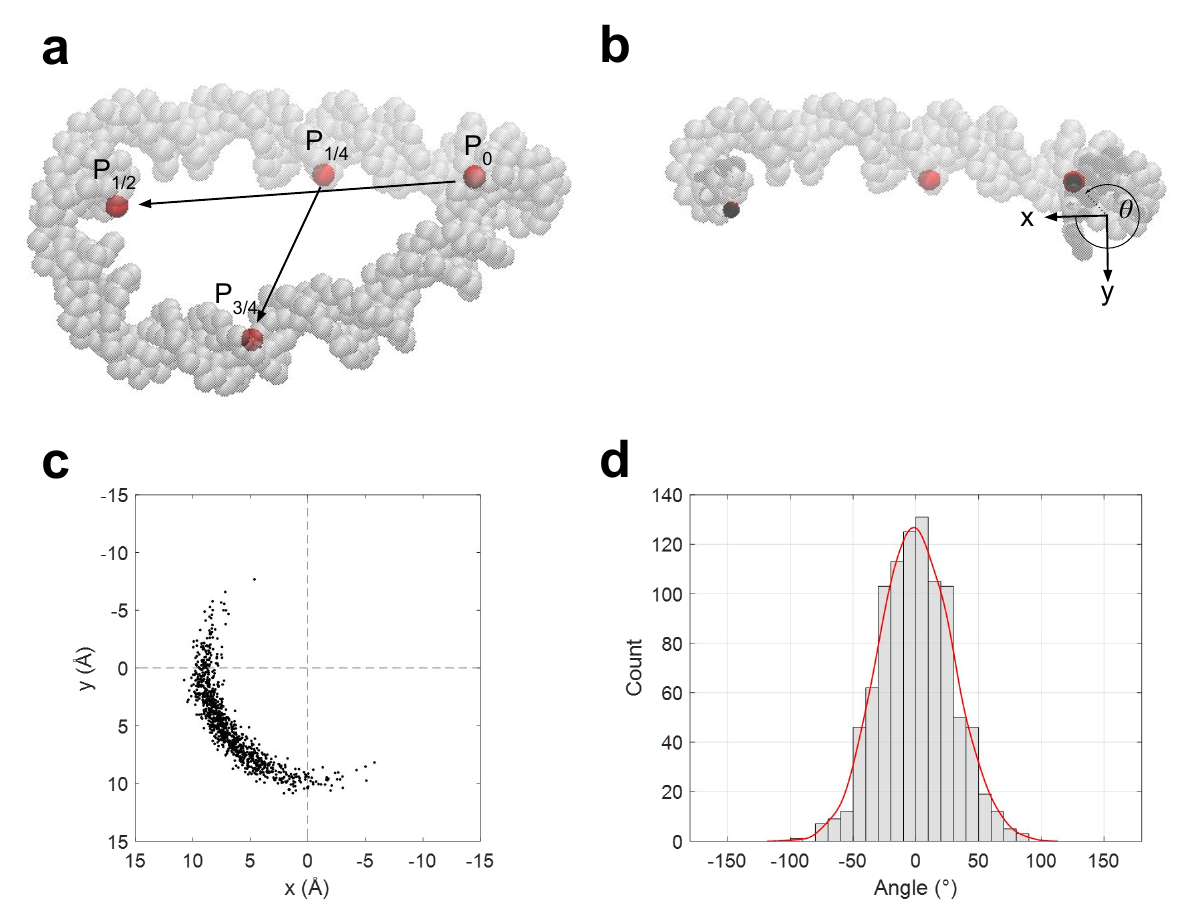}
\caption{Poloidal angle calculation using 601\_0 as an example.
(a) A representative DNA minicircle conformation from MADna simulations. Four equidistant phosphate atoms are selected along the DNA backbone, including the phosphate ($P_0$) closest to biotin-dT. The arrow from $P_0$ to $P_{1/2}$ defines the x-axis, and the arrow from $P_{1/4}$ to $P_{3/4}$ defines the z-axis. The cross product of z and x yields the y-axis.
(b) Clipped view of the same minicircle shown in (a). The xy-plane is defined as the plane containing $P_0$ and normal to the z-axis. The geometric center of mass of an 10-bp DNA segment centered at $P_0$ is computed, and its closest point on the xy-plane is designated as the origin O. The polar angle of $P_0$ with respect to this Cartesian coordinate system is defined as the poloidal angle ($\theta$).
(c) Positions of $P_0$ on the xy-plane from 950 minicircle conformations over a 95-ns simulation. The mean of the polar angles of these points is 35.9\textdegree, and the standard deviation ($\sigma$) is 28.7\textdegree. The points also vary in radial distance from the origin due to thermal fluctuations of the DNA helix.
(d) A histogram of poloidal angles, calculated from the data shown in (c). For a better visualization, the distribution is translated to have zero mean. The gray bars represent the normalized histogram obtained from 950 conformations, and the red line represents the smoothed histogram obtained via kernel density estimation. FWHM of this bell-curve-like distribution is close to $2.35 \times \sigma\approx 67$ as expected for a normal distribution.}
\label{sfig:PoloidalCalc}
\end{sfigure}
\clearpage

\begin{sfigure}[p]
\centering
\includegraphics[width=\textwidth]{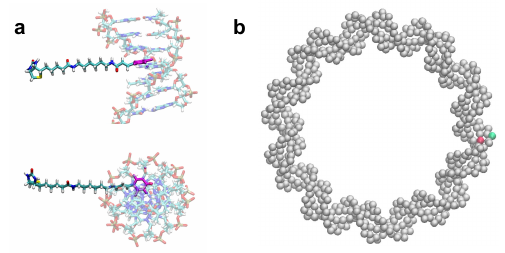}
\caption{The position of the biotin group relative to DNA. (a) A atomistic-level structural model of the biotinylated DNA segment. This model features a 7-bp DNA duplex with a biotin-modified deoxythymidine (dT) at its center. The DNA duplex is rendered transparent to highlight the biotin-dT, with the thymine base colored magenta. Using MolView \cite{bergwerf2015molview}, the biotin moiety was constructed based on the chemical structure provided by Bioneer, and its 3D geometry was optimized. The biotin was then bonded to the dT, and the bond geometry further refined using Avogadro \cite{hanwell2012avogadro}. The top panel shows a side view, illustrating the biotin emerging from the major groove of the DNA helix, while the bottom panel shows the axial view. The biotin linker is approximately as long as the helix diameter, making the biotin accessible to NeutrAvidin.
(b) Positions of phosphate and dT in a MADna-generated DNA minicircle. The red bead represents dT, and the green bead is the nearest phosphate to it. A biotin, if attached to the red bead, should emerge from the major groove, likely pointing out of the page, while the closest phosphate resides at the outermost position of the minicircle. Consequently, the poloidal angle of the phosphate is approximately 90\textdegree\space ahead of that of the biotin.}
\label{sfig:BiotinDNA}
\end{sfigure}
\clearpage

\begin{sfigure}[p]
\centering
\includegraphics[width=\textwidth]{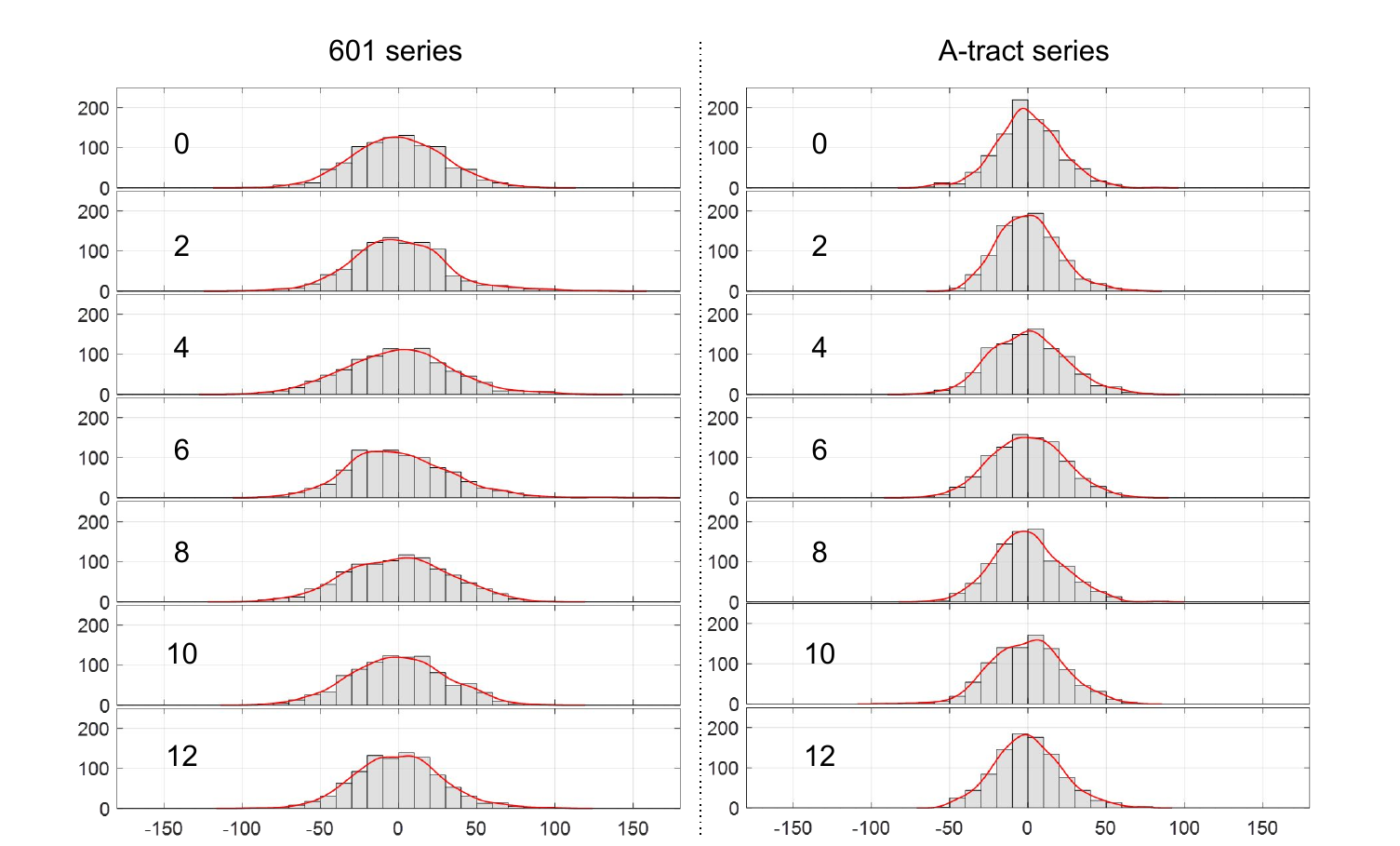}
\caption{Histograms. The left and right columns show mean-centered poloidal angle histograms from the 601 series and from the A-tract series, respectively. From the top row to the bottom, the biotin-insert position changes from 0 to 12.}
\label{sfig:histograms}
\end{sfigure}
\clearpage

\begin{sfigure}[p]
\centering
\includegraphics[width=\textwidth]{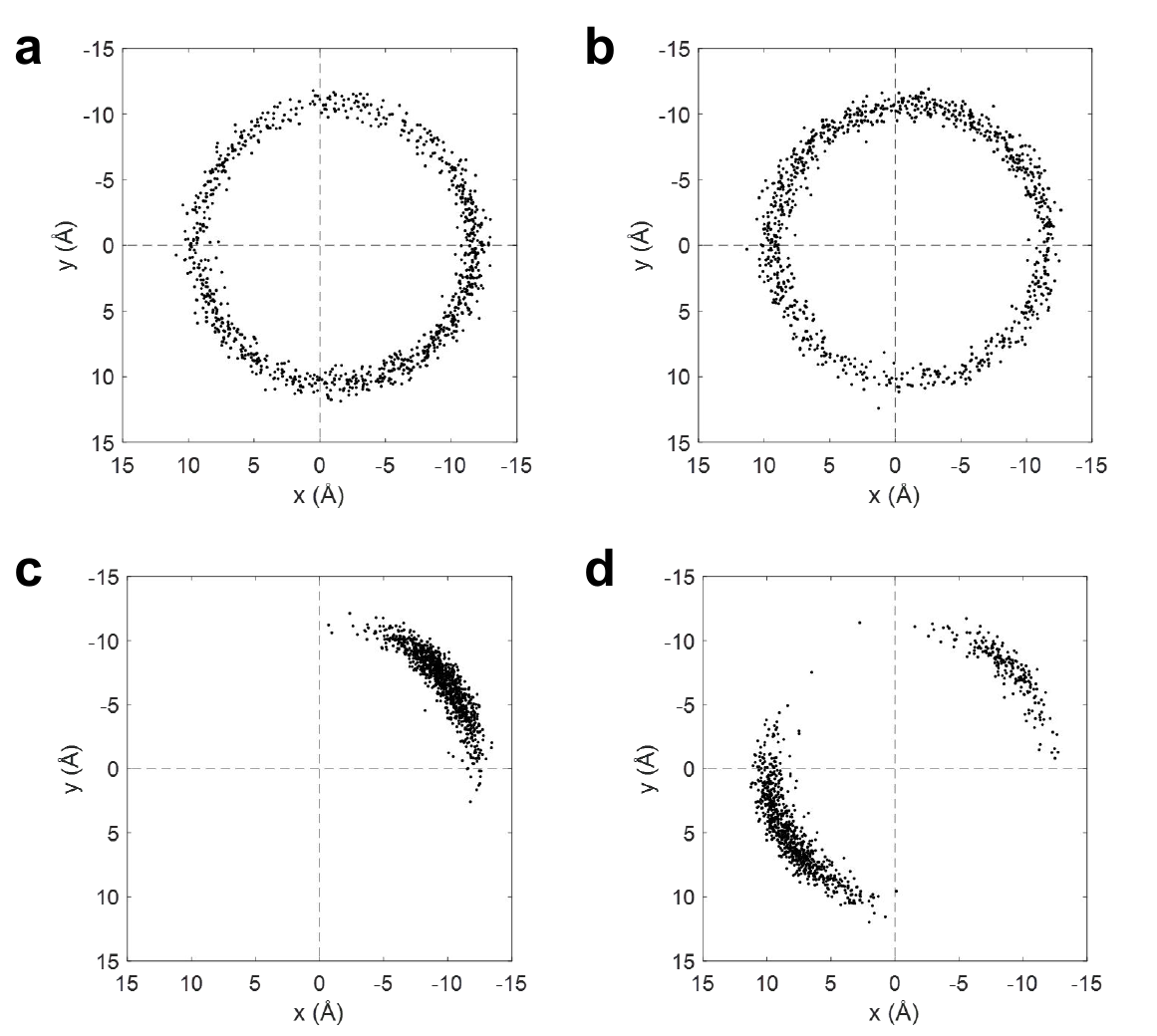}
\caption{Poloidal angle distributions of various repeat sequences. All DNA minicircles are 88 bp in length with a linking number of 8, corresponding to a helical repeat of 11 bp. The horizontal and vertical axes follow the same definition as in \autoref{sfig:PoloidalCalc}. Each dot represents the position of a phosphate atom relative to the centroid of its surrounding 10-bp DNA segment. (a) \texttt{C}\textsubscript{88}, a homopolymer of C. The geometric center of the distribution does not visibly align with the origin due to the bending of the 10-bp segment: an inward-facing phosphate atom is closer to the centroid than an outward-facing one. (b) (\texttt{TA})\textsubscript{44}, a TA dinucleotide repeat sequence (c) (\texttt{ACCCCCCCCCC})\textsubscript{8}  (d) (\texttt{ACCCCACCCCC})\textsubscript{8} }
\label{sfig:poloidaldistribution}
\end{sfigure}
\clearpage 

\clearpage

\begin{sfigure}[p]
\centering
\includegraphics[scale=.45]{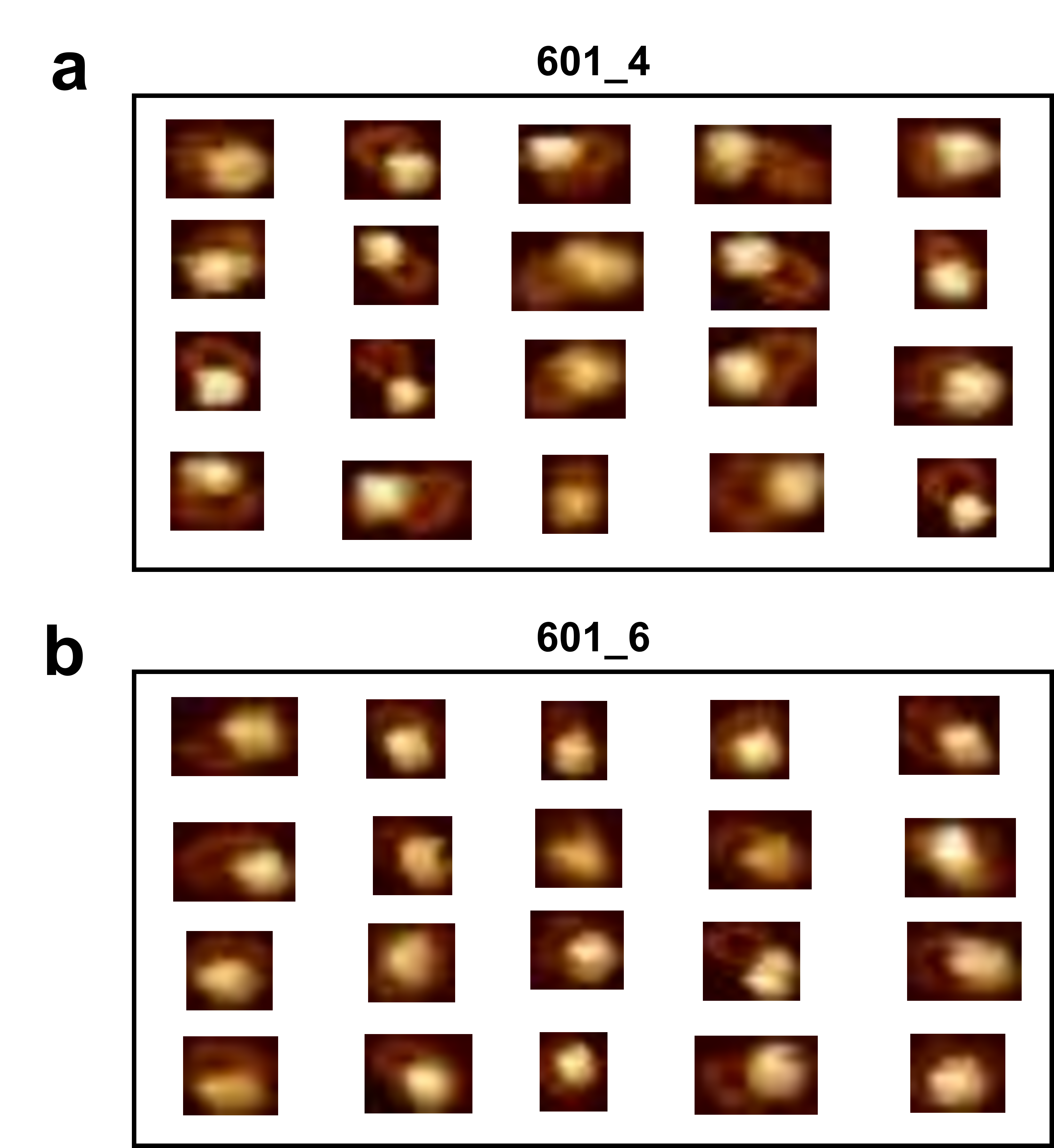}
\caption{Examples of validated DNA-protein images from 601 series for minicircles with insert position 4 (a) and 6 (b).}
\label{sfig:AFMscansSupp}
\end{sfigure}
\clearpage

\newpage
\begin{longtable}{|p{.20\linewidth}|p{.8\linewidth}| }

 \hline
 \multicolumn{2}{|c|}{\textbf{Sequence (5$'$ to 3$'$)} } \\
 \hline 
\makecell{601\_0} &
\makecell[l]{\smaller[1] \texttt{TCCGTCGAATATGATATCCTCGCGCTGTCCCCCGCGTTTTAGAATCCCGGTGCCGAGGCCGCTAAATTG} \\  \smaller[1] \texttt{GTCGTAGACAGCTCTAGCACCGCTTAAACGCACGTA} \\
         } \\ \hline
\makecell{601\_2} &
\makecell[l]{\smaller[1]
\texttt{TCCGTCGAATATGATATCCTTACGCGCTGTCCCCCGCGTTTTAGAATCCCGGTGCCGAGGCCGCTAAAT} \\  \smaller[1] \texttt{TGGTCGTAGACAGCTCTAGCACCGCTTAAACGCACG} \\
         } \\ \hline
\makecell{601\_4} &
\makecell[l]{\smaller[1] \texttt{TCCGTCGAATATGATATCCTCGTACGCGCTGTCCCCCGCGTTTTAGAATCCCGGTGCCGAGGCCGCTAA} \\   \smaller[1] \texttt{ATTGGTCGTAGACAGCTCTAGCACCGCTTAAACGCA} \\
        } \\ \hline
\makecell{601\_6} &
\makecell[l]{\smaller[1] \texttt{TCCGTCGAATATGATATCCTCACGTACGCGCTGTCCCCCGCGTTTTAGAATCCCGGTGCCGAGGCCGCT} \\    \smaller[1] \texttt{AAATTGGTCGTAGACAGCTCTAGCACCGCTTAAACG} \\
         } \\ \hline
\makecell{601\_8} &
\makecell[l]{\smaller[1] \texttt{TCCGTCGAATATGATATCCTCGCACGTACGCGCTGTCCCCCGCGTTTTAGAATCCCGGTGCCGAGGCCG} \\    \smaller[1] \texttt{CTAAATTGGTCGTAGACAGCTCTAGCACCGCTTAAA} \\
        } \\ \hline
\makecell{601\_10} &
\makecell[l]{\smaller[1] \texttt{TCCGTCGAATATGATATCCTAACGCACGTACGCGCTGTCCCCCGCGTTTTAGAATCCCGGTGCCGAGGC} \\  \smaller[1] \texttt{CGCTAAATTGGTCGTAGACAGCTCTAGCACCGCTTA} \\
        } \\ \hline
\makecell{601\_12} &
\makecell[l]{\smaller[1] \texttt{TCCGTCGAATATGATATCCTTAAACGCACGTACGCGCTGTCCCCCGCGTTTTAGAATCCCGGTGCCGAG} \\  \smaller[1] \texttt{GCCGCTAAATTGGTCGTAGACAGCTCTAGCACCGCT} \\
        } \\ \hline
\makecell{A-tract\_0}&\makecell[l]{\smaller[1] \texttt{TCCGTCGAATATGATATCCTCGAAAAAACGGGCAAAAAACGGCAAAAAACGGGCAAAAAACGGCAAAAA}\\
\smaller[1] \texttt{ACGGGCAAAAAATCTAGCACCGCTTAAACGCACGTA}                                  }       \\ \hline
\makecell{A-tract\_2}&\makecell[l]{\smaller[1] \texttt{TCCGTCGAATATGATATCCTTACGAAAAAACGGGCAAAAAACGGCAAAAAACGGGCAAAAAACGGCAAA}\\
\smaller[1] \texttt{AAACGGGCAAAAAATCTAGCACCGCTTAAACGCACG}\\
        } \\ \hline 
\makecell{A-tract\_4}&\makecell[l]{\smaller[1]  \texttt{TCCGTCGAATATGATATCCTCGTACGAAAAAACGGGCAAAAAACGGCAAAAAACGGGCAAAAAACGGCA}\\
 \smaller[1] \texttt{AAAAACGGGCAAAAAATCTAGCACCGCTTAAACGCA}     \\
        } \\ \hline 
\makecell{A-tract\_6}&\makecell[l]{\smaller[1] \texttt{TCCGTCGAATATGATATCCTCACGTACGAAAAAACGGGCAAAAAACGGCAAAAAACGGGCAAAAAACGG}\\
\smaller[1]   \texttt{CAAAAAACGGGCAAAAAATCTAGCACCGCTTAAACG}\\
        } \\ \hline 
\makecell{A-tract\_8}&\makecell[l]{\smaller[1] \texttt{TCCGTCGAATATGATATCCTCGCACGTACGAAAAAACGGGCAAAAAACGGCAAAAAACGGGCAAAAAAC}\\
\smaller[1] \texttt{GGCAAAAAACGGGCAAAAAATCTAGCACCGCTTAAA}\\
        } \\ \hline 
\makecell{A-tract\_10}&\makecell[l]{\smaller[1] \texttt{TCCGTCGAATATGATATCCTAACGCACGTACGAAAAAACGGGCAAAAAACGGCAAAAAACGGGCAAAAA}\\
\smaller[1] \texttt{ACGGCAAAAAACGGGCAAAAAATCTAGCACCGCTTA}\\
       }  \\ \hline 
\makecell{A-tract\_12}&\makecell[l]{\smaller[1] \texttt{TCCGTCGAATATGATATCCTTAAACGCACGTACGAAAAAACGGGCAAAAAACGGCAAAAAACGGGCAAA}\\
\smaller[1] \texttt{AAACGGCAAAAAACGGGCAAAAAATCTAGCACCGCT}\\
        }  \\ \hline   \hline
                 
 \multicolumn{2}{|c|}{\textbf{Primers for making minicircles (5$'$ to 3$'$)}} \\\hline 

    \makecell{Forward}& \texttt{[Phos]TCCGTCGAA[biotindT]ATGATATCCT}  \\\hline
    \makecell{601\_0 reverse}&  \texttt{[Phos]TACGTGCGTTTAAGCGGT} \\\hline 
    \makecell{601\_2 reverse}&  \texttt{[Phos]CGTGCGTTTAAGCGGTG}\\\hline
    \makecell{601\_4 reverse}&  \texttt{[Phos]TGCGTTTAAGCGGTGCTA} \\\hline 
    \makecell{601\_6 reverse}&  \texttt{[Phos]CGTTTAAGCGGTGCTAGA} \\\hline 
    \makecell{601\_8 reverse}&  \texttt{[Phos]TTTAAGCGGTGCTAGAGC} \\\hline 
    \makecell{601\_10 reverse}& \texttt{[Phos]TAAGCGGTGCTAGAGCTG} \\\hline 
    \makecell{601\_12 reverse}& \texttt{[Phos]AGCGGTGCTAGAGCTGTC} \\\hline 

   \makecell{A-tract\_0 reverse}& \texttt{[Phos]TACGTGCGTTTAAGCGGT} \\\hline 
   \makecell{A-tract\_2 reverse}& \texttt{[Phos]CGTGCGTTTAAGCGGTG} \\\hline
   \makecell{A-tract\_4 reverse}& \texttt{[Phos]TGCGTTTAAGCGGTGCTA} \\\hline 
   \makecell{A-tract\_6 reverse}& \texttt{[Phos]CGTTTAAGCGGTGCTAGA} \\\hline 
   \makecell{A-tract\_8 reverse}& \texttt{[Phos]TTTAAGCGGTGCTAGATT} \\\hline 
   \makecell{A-tract\_10 reverse}& \texttt{[Phos]TAAGCGGTGCTAGATTTT} \\\hline 
   \makecell{A-tract\_12 reverse}& \texttt{[Phos]AGCGGTGCTAGATTTTTT} \\\hline

 \caption{List of DNA sequences and PCR primers for creating minicircles for AFM experiments.}
\label{tab:DNA_SequencesAFM}\\ 
                           

\end{longtable}
\clearpage

\begin{table}[h]
\centering
\begin{tabular}{|c|c|c|c|}
\hline
Atom & $r (\text{\AA})$ & $\theta_0 (^\circ)$& $z_0 (\text{\AA})$ \\
\hline
S & 8.7 & 0 & 0 \\
B & 4.2 & 5.1 & -0.2 \\
P & 10.6 & 10.2 & 2.7 \\
S2 & 8.7 & 102.4 & -1.2 \\
B2 & 4.0 & 93.6 & -1.0 \\
P2 & 10.6 & 123.8 & -0.6 \\
\hline
\end{tabular}
\caption{Parameters used to construct DNA for MADna simulations.}
\label{tab:DNAinitial}
\end{table}
\clearpage

\newpage
\begin{table}[h]
\centering
\begin{tabular}{|l|r|r|r|r|}
\hline
\textbf{Name} & \textbf{Mean angle (°)} & \textbf{Std (°)} & \textbf{Mean x (Å)} & \textbf{Std (Å)} \\
\hline
601\_0       & 35.9   & 28.7   & 6.5   & 2.7 \\
601\_2       & 93.7   & 30.7   & -0.7  & 4.5 \\
601\_4       & -160.9 & 34.1   & -8.5  & 3.0 \\
601\_6       & -99.9  & 32.5   & -1.7  & 4.8 \\
601\_8       & -19.7  & 32.4   & 7.3   & 2.2 \\
601\_10      & 25.6   & 30.3   & 7.1   & 2.3 \\
601\_12      & 78.5   & 28.2   & 1.5   & 4.1 \\
A-tract\_0   & -171.1 & 20.6   & -10.2 & 1.3 \\
A-tract\_2   & -112.5 & 19.2   & -3.9  & 3.2 \\
A-tract\_4   & -45.7  & 23.8   & 5.9   & 2.5 \\
A-tract\_6   & 24.3   & 23.2   & 7.6   & 1.7 \\
A-tract\_8   & 88.7   & 21.9   & 0.1   & 3.5 \\
A-tract\_10  & -176.9 & 22.7   & -10.2 & 1.4 \\
A-tract\_12  & -120.6 & 21.2   & -5.2  & 3.4 \\
\hline
\end{tabular}
\caption{Mean poloidal angle and $x$ position with standard deviations for each DNA minicircle.}
\label{tab:angle_x_stats}
\end{table}
\end{document}